\documentclass[11pt]{article}

\usepackage{amsmath, amssymb, amsfonts}
\usepackage{geometry}
\usepackage{hyperref}
\usepackage{graphicx}
\usepackage{bm}
\usepackage{float}
\title{Topologically Stabilized Torsion in Weak-Field Gravity: A Ricci--Flow Framework}
\author{Elisa Varani \\
\small Universit\`a Cattolica del Sacro Cuore \\
\small Milano, Italia \\
\small \texttt{elisa.varani@unicatt.it}}
\date{}

\begin{document}
\maketitle

\begin{abstract}
We investigate stationary torsional configurations supported by chiral Majorana
neutrino currents in linearized gravity. A Ricci--flow--inspired geometric relaxation
(with no physical time interpretation) is introduced to drive the metric
perturbation toward fixed points sustained by chiral sources while keeping
curvature invariants negligible. We show that divergence-free chiral currents can
support globally non-trivial torsional holonomy stabilized by topological
invariants associated with $\pi_1(S^1)$ and $\pi_3(S^3)$. Toroidal skyrmionic
domains emerge when one chirality dominates, whereas a chiral-flip interference
sector enables M\"obius-type non-orientable bridges between opposite-chirality
regions. In the static limit, a Green-function formulation provides a finite-range
Yukawa-type response governed by the neutrino coherence length. These results
identify a purely torsional mechanism --- independent of local curvature --- through
which coherent chiral currents may influence effective gravitational behavior in
neutrino-rich environments.
\end{abstract}


\section{Introduction}
Spinor fields in gravitational theories with torsion couple to the antisymmetric part of the affine connection~\cite{Hehl,Shapiro}.
Majorana neutrinos, with intrinsic chirality, are natural sources for torsional effects and may form coherent domains.
This short communication focuses on \emph{stationary} torsional solutions supported by chiral currents, which are topologically non--trivial while the curvature remains negligible. Coherent neutrino populations are expected to arise in several
high-density astrophysical environments — from supernova interiors
to the early Universe neutrino background — making torsion-induced
effects of potential relevance beyond laboratory scales.

A broader development, including extended classification and topological aspects, appears in a companion study~\cite{Varani2025}.
Related Yukawa--torsion effects are discussed in~\cite{Varani}.

\subsection{Chiral currents and weak--field setting}

We consider the weak--field expansion of the metric,
\begin{equation}
g_{\mu\nu} = \eta_{\mu\nu} + h_{\mu\nu}, 
\qquad |h_{\mu\nu}|\ll 1,
\end{equation}
and focus on Majorana spinor fields, whose left- and right-handed components are
related by charge conjugation. Their intrinsic chirality makes them natural sources
for torsional degrees of freedom.

The left- and right-handed fermionic currents are defined as
\begin{equation}
J^\mu_{L} = \bar\psi \gamma^\mu P_L \psi, \qquad
J^\mu_{R} = \bar\psi \gamma^\mu P_R \psi, \qquad
P_{L,R} = \frac{1\mp\gamma_5}{2},
\end{equation}
and encode the local transport of chiral charge. In spatial components,
\begin{equation}
J^k_{L,R} = \bar\psi_{L,R}\,\sigma^k\,\psi_{L,R},
\end{equation}
with $\sigma^k$ the Pauli matrices.

Torsion is sourced by the chiral imbalance between the left- and right-handed
sectors, schematically $(J_L - J_R)$, and by an interference (flip) sector built from
mixed bilinears of the form $\bar\psi_L\psi_R$. The latter enables the formation of
globally non-orientable torsional structures despite locally vanishing curvature.

\subsection{Ricci flow evolution}

To model the geometric response of the torsional sector to chiral currents, we use a
flow parameter $\tau$ that evolves the metric perturbation toward stationary
configurations without representing physical time. 
In the weak--field regime, the evolution equation takes the form
\begin{equation}
\frac{\partial h_{\mu\nu}}{\partial\tau}
= -2\,R_{\mu\nu}(h),
\end{equation}
inspired by the classical Ricci flow framework introduced by Hamilton and
extended to include entropy functionals by Perelman~\cite{Hamilton,Perelman}.
Here $R_{\mu\nu}(h)$ is the linearized Ricci tensor and encodes the
torsional contribution generated by chiral currents~\cite{Varani2025}.

Using the Levi--Civita structure $\epsilon_{abc}$, the most relevant components can be
expressed schematically as
\begin{align}
\frac{\partial h_{0c}}{\partial \tau}
&\propto
\int dx^b\,\epsilon_{bck}
\left(\bar\psi_L\sigma^k\psi_L
- \bar\psi_R\sigma^k\psi_R\right),
\label{flow1}
\\
\frac{\partial h_{ac}}{\partial \tau}
&\propto
\int dx^0\,\epsilon_{ack}
\left(\bar\psi_L\sigma^k\psi_L
- \bar\psi_R\sigma^k\psi_R\right),
\label{flow2}
\\
\left.
\frac{\partial h_{ac}}{\partial \tau}
\right|_{\rm flip}
&\propto
\int dx^b\,\epsilon_{bca}
\left(\bar\psi_L\psi_R
- \bar\psi_R\psi_L\right),
\label{flow3}
\end{align}
demonstrating that left- and right-handed currents induce torsion with opposite sign,
while the flip bilinear couples the two chiral components and enables the formation
of globally coherent, possibly non-orientable configurations.

The aim of the flow evolution is to identify geometry--current
combinations that satisfy $\partial_\tau h_{\mu\nu}=0$, i.e.\ stationary torsional states,
which are analyzed in the next section.

\subsection{Left/right and chiral-flip contributions}

The Levi--Civita tensor $\epsilon_{abc}$ in Eqs.~(\ref{flow1})--(\ref{flow3}) enforces
an opposite orientation of torsion generated by left- and right-handed currents.
Domains dominated by $J_L$ and $J_R$ therefore induce counter-rotating local frames,
realizing geometrically distinct chiral phases with vanishing curvature.

The flip sector, driven by mixed bilinears of the form $\bar{\psi}_L\psi_R$, couples
the two chiral components and allows for a continuous interpolation between them.
This mechanism acts as a torsional ``stitching'' that preserves stationarity while
modifying the global orientability of the configuration.

\begin{figure}[H]
\centering
\fbox{\includegraphics[width=0.4\linewidth]{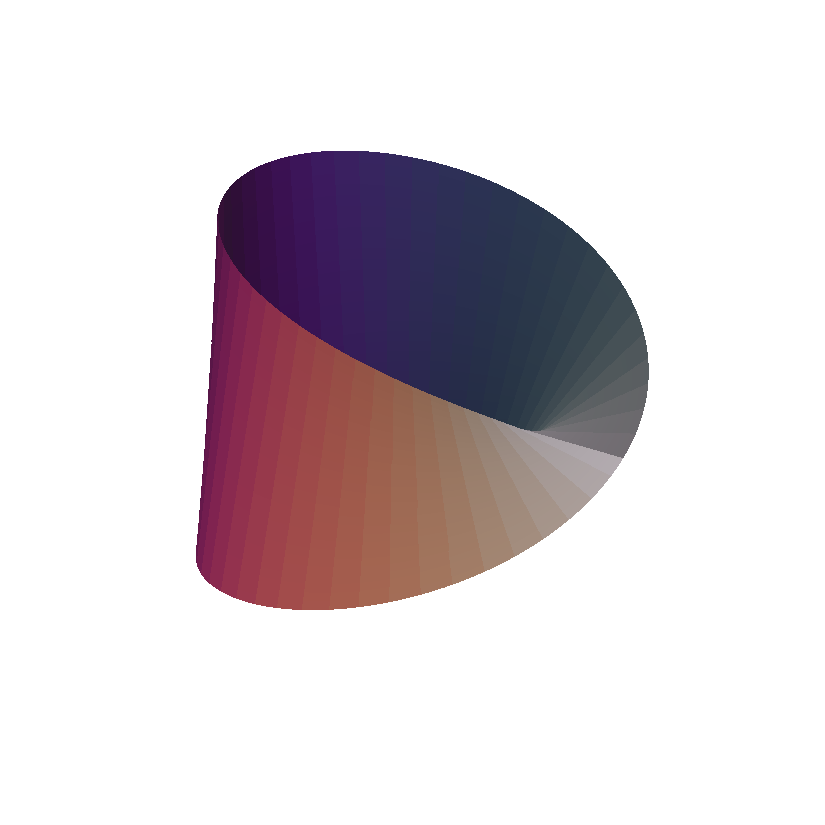}}
\caption{Schematic M\"obius-like torsional bridge generated by chiral-flip stitching between
domains of opposite chirality. Local curvature remains negligible while the global
holonomy is non-trivial.}
\label{fig:mobius}
\end{figure}

Coherent domains connected through the flip sector can thus produce globally
non-trivial torsional structures --- such as toroidal skyrmionic regions or
M\"obius-like bridges --- where all curvature invariants remain arbitrarily small.
These configurations will be classified topologically in the next section.

\section{Topological characterization}

Stationary torsional configurations fall into distinct topological classes depending
on whether torsion is confined along a closed line or distributed over a tubular
domain. These properties are captured by global invariants associated with the
holonomy of the torsional connection and its mapping into an internal chiral space.
\subsection{Holonomy and torsional vortices}

The torsional sector can be represented by a gravitational vector potential
$A^g_\mu$ ~\cite{Varani2025}, such that parallel transport acquires a phase depending solely on the
global structure of the connection. The torsional circulation along a closed
loop $\Gamma$ is quantified by the holonomy:
\begin{equation}
\Phi_T = \oint_\Gamma A^g_\mu \, dx^\mu = 2\pi n, 
\qquad n \in \mathbb{Z} \, .
\end{equation}
A non-zero integer $n$ signals the presence of a line-like torsional vortex,
classified by the first homotopy group $\pi_1(S^1)$. The spacetime remains locally flat but exhibits a
globally non-trivial phase --- a torsional analogue of the Aharonov--Bohm
effect in gauge theory. Thus, $\Phi_T$ detects the existence of a quantized
torsional defect.

\subsection{Skyrmionic torsion and toroidal domains}

Beyond line-like defects, torsion can also form volumetric domains where the
orientation of the chiral frame winds non-trivially throughout a 3D region. This
is characterized by a unit vector field $\hat{\mathbf{n}}(\mathbf{x})$ associated with the
local chiral frame, whose global wrapping defines the topological charge:
\begin{equation}
Q_S = \int d^3x \
\epsilon^{ijk} \
(\partial_i \hat{\mathbf{n}})\cdot
\big(\partial_j \hat{\mathbf{n}} \times \partial_k \hat{\mathbf{n}}\big)
\in \mathbb{Z} \, .
\end{equation}
A non-zero $Q_S$ corresponds to a $\pi_3(S^3)$ classification, ensuring that the
torsional domain cannot be continuously unwound without destroying or severing
the region that supports it. In this sense, $Q_S \neq 0$ stabilizes toroidal
skyrmion-like structures as stationary solutions independent of the holonomy
associated with $\Phi_T$.

In summary, $\Phi_T$ identifies a torsional vortex and $Q_S$ guarantees that it
remains topologically protected: the former detects the defect, the latter
prevents it from disappearing.

\section{Examples of stationary torsional domains}

Stationary solutions supported by chiral currents naturally organize into coherent
domains whose topology is fixed by the structure of the currents. Here we present two
representative geometries, corresponding to the two distinct topological invariants
introduced in the previous section: a toroidal skyrmionic configuration with non-zero
$\pi_3(S^3)$ charge, and a M\"obius-like torsional bridge associated with a 
$\pi_1(S^1)$ holonomy induced by the flip sector.

\subsection{Toroidal stationary domain (skyrmion-like)}

A domain in which left-handed currents circulate around a closed loop yields a
tubular region with non-trivial $\pi_3(S^3)$ charge. The torsion remains localized
within the torus while curvature stays negligible everywhere. The internal circulation
encodes the skyrmionic structure and stabilizes the configuration against smooth
deformations.

\begin{figure}[H]
\centering
\fbox{\includegraphics[width=0.4\linewidth]{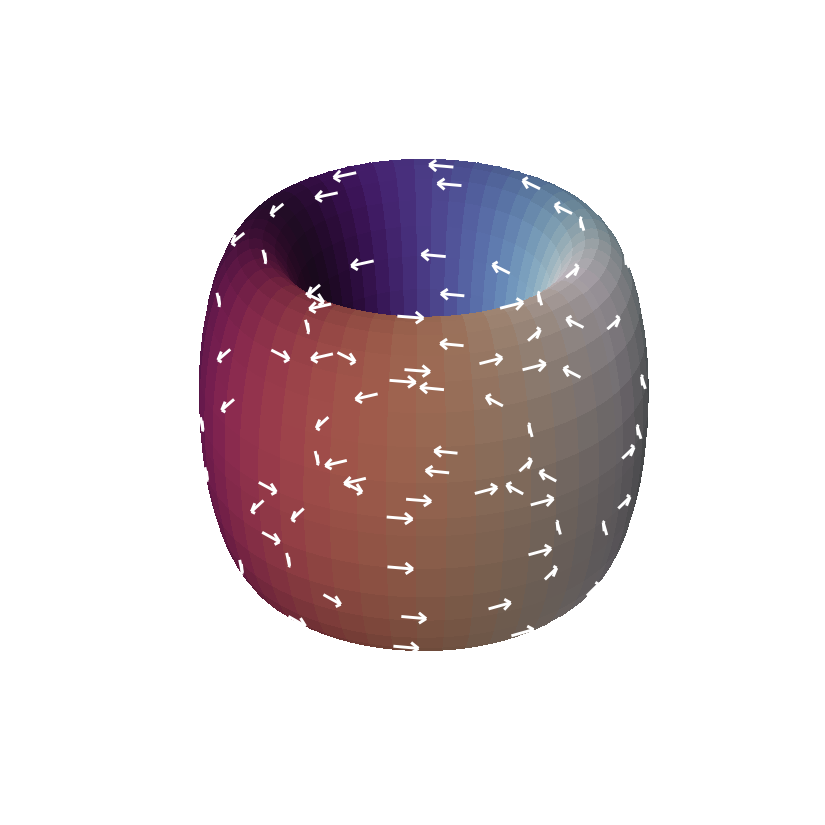}}
\caption{Schematic toroidal torsional domain supported by circulating chiral currents.
The torsional geometry is globally stable due to a non-zero skyrmion charge.}
\label{fig:toroid}
\end{figure}

\subsection{Chiral-flip stitching: M\"obius-like bridge}

When two regions dominated by opposite chirality are connected through the flip
sector, the resulting torsional field may become globally non-orientable while
remaining flat locally. The holonomy around the strip leads to a non-trivial
$\pi_1(S^1)$ classification, as in Aharonov--Bohm-type defects.

These structures demonstrate that topologically non-trivial torsion can be
sustained solely by chiral imbalance, providing explicit realizations of the
stationary solutions described above.

\section{Static limit and Green-function formulation}

The stationary condition in the flow parameter $\tau$ does not necessarily imply the absence
of residual spatial variations. It is thus useful to complement the Ricci--flow--based
description with a static limit, where $\partial_{0}h_{\mu\nu}=0$ and time derivatives vanish
in physical coordinates. In this regime, the linearized field equations reduce to an elliptic
system
\begin{equation}
\nabla^{2}h_{\mu\nu}(\bm{x})=-2\,T^{\rm eff}_{\mu\nu}(\bm{x}),
\end{equation}
where $T^{\rm eff}_{\mu\nu}$ is built directly from localized chiral sources (including the
flip interference). This formulation highlights the finite-range response of the torsional
sector to the underlying spinor currents.

A screened Green function,
\begin{equation}
G(\bm{x},\bm{x}')=\frac{e^{-\mu|\bm{x}-\bm{x}'|}}{4\pi|\bm{x}-\bm{x}'|},
\end{equation}
accounts for a characteristic coherence length $\mu^{-1}$ of the torsion-inducing neutrino
currents. The metric perturbation can then be written in compact Yukawa-type form:
\begin{align}
h_{0c}(\bm{x})&=\frac{1}{3}\int d^{3}x'\,
G(\bm{x},\bm{x}')\,\epsilon_{bck}
\big(\bar\psi_L\sigma^k\psi_L-\bar\psi_R\sigma^k\psi_R\big)(\bm{x}'),
\\
h_{ac}^{(L/R)}(\bm{x})&=-\frac{1}{3}\int d^{3}x'\,
G(\bm{x},\bm{x}')\,\epsilon_{ack}
\big(\bar\psi_L\sigma^k\psi_L-\bar\psi_R\sigma^k\psi_R\big)(\bm{x}'),
\\
h_{ac}^{(\rm flip)}(\bm{x})&=-\frac{1}{3}\int d^{3}x'\,
G(\bm{x},\bm{x}')\,\epsilon_{bca}
\big(\bar\psi_L\psi_R-\bar\psi_R\psi_L\big)(\bm{x}')\,.
\end{align}
These expressions show that left- and right-handed contributions retain opposite sign,
while the flip sector couples chiral domains, thus enabling globally non-orientable
stationary configurations such as those depicted in Figs.~\ref{fig:toroid}--\ref{fig:mobius}.
Complementary Yukawa–torsion analysis is developed by the author in ~\cite{Varani}.

\section{Conclusion}

We have shown that chiral Majorana neutrino currents can sustain stationary torsional
structures in the weak-field regime through a Ricci--flow--inspired evolution of the metric
perturbation. These configurations include toroidal domains, stabilized by a skyrmion
charge associated with $\pi_3(S^3)$, and M\"obius-like bridges where a chiral-flip sector
induces a non-orientable global geometry while local curvature remains negligible. The
resulting states are characterized by conserved topological invariants (vortex flux and
skyrmion number) and a finite-range response encoded by a Yukawa-type Green function. 
The screened Green–function used here is derived and discussed 
in~\cite{Varani}.

Such torsional stationary domains could, in principle, emerge on scales where coherent
neutrino currents are present, potentially influencing the effective gravitational behavior
of regions that do not contain ordinary matter. Left-handed configurations may contribute
to localized attractive geometric effects, while right-handed sectors provide an opposite
response. A network of such stationary torsional structures may therefore play a role in
large-scale geometry, possibly contributing to gravitational effects that are not directly
associated with visible matter.

A detailed investigation of these broader implications, including their potential
relevance in astrophysical and cosmological contexts, is developed in the extended
companion study~\cite{Varani2025}.

\section*{Acknowledgements}
The author acknowledges Francisco Bulnes for helpful discussions that contributed to this work.

\end{document}